\documentclass[11pt,twoside]{article}

\usepackage{asp2006}
\usepackage{natbib}
\usepackage{graphicx}

\newcommand{\msun}{M$_\odot$}
\newcommand{\lsun}{L$_\odot$}
\newcommand{\mum}{$\mu$m}
\newcommand{\degree}{$^o$}

\newcommand{\vlsr}{v$_{\rm{LSR}}$}

\begin{document}
\title{The RMS Survey: A Galaxy-wide Sample of Massive Young Stellar Objects}   %%% Fill in title
\author{J.~S.~Urquhart$^1$, M.~G.~Hoare$^1$, S.~L.~Lumsden$^1$, R.~D.~Oudmaijer$^1$, T.~J.~T.~Moore$^2$}   %%% Fill in author names
\affil{$^1$School of Physics and Astrophysics, University of Leeds, Leeds, LS2~9JT, UK, Astrophysics Research Institute\\ $^2$Liverpool John Moores University, Twelve Quays House, Egerton Wharf, Birkenhead, CH41~1LD, UK}    %%% Fill in author affiliations

\begin{abstract} %%% Abstract to run on from here.

Here we describe the Red MSX Source (RMS) survey which is the
largest, systematic, galaxy-wide search for massive young stellar
objects (MYSOs) yet undertaken. Mid-IR bright point sources from the
MSX satellite survey have been followed-up with ground-based radio,
millimetre, and infrared observations to identify the contaminating
sources and characterise the MYSOs and UCHII regions. With the initial
classification now complete the distribution of sources in the
galaxy will be discussed, as well as some programmes being developed
to exploit our sample.

\end{abstract}

%%% MAIN BODY OF TEXT GOES HERE. CONSULT "INSTRUCTIONS FOR AUTHORS USING
%%% LATEX2E MARKUP", SECTIONS 2.3-2.6 FOR HELP WITH EQUATIONS, FIGURES,
%%% AND TABLES.

%\section{}   %%% Top level section head (remove "%" symbol)
%\subsection{}   %%% Second level section head (remove "%" symbol)
%\subsubsection{}   %%% Lowest level section head (remove "%" symbol)
%\section*{}    %%% Unnumbered top level section head (remove "%" symbol)
%\subsection*{}   %%% Unnumbered second level section head (remove "%" symbol)

\section{Introduction}

Massive stars ($M$$>$8~\msun, $L$$>$10$^4$~\lsun) play a fundamental role in
many areas of astrophysics. They are the principal source of UV
radiation and heavy elements in galaxies, and are responsible for
injecting huge amounts of kinetic energy into the ISM through powerful
molecular outflows, strong stellar winds and supernova explosions. The
momentum imparted through these processes provides an important source
of mixing and turbulence within the ISM. They are also thought to play
a key role in regulating star formation, either by disrupting
molecular clouds before stars have been able to form, or
constructively through the expansion of their HII regions (e.g.,
collect and collapse) or the propagation of strong shocks into their
surroundings (e.g., radiatively driven implosion). Massive stars have an
enormous influence not only on the physical structure and chemistry of
their local environment, but also on the structure and evolution of
their host galaxies.

Despite the importance of massive stars the processes involved in
their formation and the early stages of their evolution are still
poorly understood. Massive stars reach the main sequence while still
deeply embedded within their natal molecular clouds, and therefore
their entire pre-main sequence formation and evolution is hidden
behind high levels of extinction. They are extremely rare and as a
consequence are generally located farther away than regions of
low-mass star formation. The large distances are compounded by the
fact that massive stars are through to form exclusively in clusters
and limited spatial resolution makes it difficult to identify and
attribute derived quantities to individual sources. Furthermore, the
evolution of massive stars is extremely rapid, which means that key
stages in their early evolution are short lived. Due to these
observational difficulties, until relatively recently, the only
catalogue of massive young stellar objects (MYSOs) had been limited to
30 or so serendipitously detected sources (\citealt{henning1984}) most of 
which are nearby.

\section{Colour-selection of MYSO candidates}

As massive stars reach the main sequence while still deeply embedded,
the obvious place to begin to search for MYSOs is at infrared wavelengths where the majority of the
bolometric luminosity ($>$10$^4$ L$_\odot$;
\citealt{wynn-williams1982}) is emitted after reprocessing by
dust. Nuclear fusion has almost certainly begun due to the very short
Kelvin-Helmholtz contraction time scale compared to their free-fall
time scale (\citealt{behrend2001}). Accretion is likely to be taking
place as the majority of MYSOs are associated with massive outflows
(\citealt{wu2004}), generally thought to be powered by
accretion. Although fusion is taking place, the ongoing accretion is
thought to impede the formation of an HII region either by: 1)
significantly increasing the star's radius, leading to a lower
effective stellar temperature and characteristic accretion disk
temperature (\citealt{hoarefranco2007}), or 2) the HII region being
quenched by high infall rates (\citealt{walmsley1995}). MYSOs are also
known to possess ionised stellar winds that are weak thermal radio
sources ($\sim$1~mJy at 1 kpc; \citealt{hoare2002}). MYSOs can be
roughly parametrised as mid-infrared bright, and thus slightly later
than the hot core phase, and radio quiet, and thus earlier
than the UCHII region phase, with luminosities consistent with young O
and early B-type stars.

To date there have been several attempts to search for
MYSOs using colour-selection criteria and the IRAS point source
catalogue (e.g.,
\citealt{molinari1996,walsh1997,sridharan2002}). However, due to confusion in the large IRAS beam ($\sim$3-5\arcmin\ at 100 $\mu$m) these samples
tended to avoid dense clustered environments and the Galactic
mid-plane where the majority of MYSOs are expected to be located since
the scale height of massive stars is $\sim$30\arcmin\
(\citealt{reed2000}). Although these IRAS selected samples have
identified many genuine MYSOs, they tend to be biased towards bright,
isolated sources that are probably not representative of the general
population of MYSOs. There is a need for an unbiased
sample of MYSOs that is large enough to have a sufficient number of
sources in each luminosity bin to allow the processes involved in
massive star formation to be investigated in a statistically robust
way.

The Midcourse Space eXperiment (MSX; \citealt{price2001}) surveyed the
whole Galactic plane ($|b|$$<$5\degree) in four mid-infrared bands
centred at 8, 12, 14 and 21~\mum\ at a resolution of 18\arcsec\ in all
bands. Although the sensitivity of MSX is similar to that of IRAS, its
beam size is $\sim$50 times smaller than that of the IRAS 12 and
25~\mum\ bands and thus avoids most of the confusion problem found with
IRAS. The MSX survey therefore offers us the opportunity, for the
first time, to identify a truly representative sample of MYSOs located
throughout the Galaxy. More recently the Spitzer satellite has
surveyed a large region of the inner Galaxy (GLIMPSE;
\citealt{benjamin2005}) with unprecedented spatial resolution and
sensitivity, offering a great improvement on MSX data. However, most
MYSOs are found to be highly saturated and would be missed in any
GLIMPSE colour-selected sample.

In order to select potential MYSOs we compared the colours of known MYSOs with  sources
identified in the MSX and 2MASS point source catalogues
(\citealt{egan2003} and \citealt{cutri2003} respectively) and
developed colour selection criteria (\citealt{lumsden2002}). A lower
limit of 2.7~Jy was placed on the 21~\mum\ flux. The
detection of a source in 2MASS was not necessary as MYSOs can be totally
obscured in the near-infrared; these data were just used to exclude
sources that are too blue. The MSX images of all colour selected sources were
visually inspected to eliminate sources that are not
point-like (see below). Sources toward the Galactic centre  ($|l|<10$\degree) were excluded
 due to confusion and difficulties in calculating
kinematic distances. In total we identified approximately 2000 MYSO
candidates.

\section{Multi-wavelength follow-up observations}

The shape of the spectral energy distribution from an optically thick
cloud is insensitive to the type of heating source. This is a problem
with a colour-selected sample such as ours, as it results in contamination
of our sample by several other types of embedded, or dust enshrouded
objects that have similar near- and mid-infrared colours to MYSOs (e.g., ultra
compact (UC) HII regions, evolved stars and planetary nebulae
(PNe)). The core of the RMS survey is a multi-wavelength
programme of follow-up observations designed to distinguish between
genuine MYSOs and these other types of embedded or dusty objects
(\citealt{hoare2005}) and to compile a database of complementary
multi-wavelength data with which to study their
properties.

The main source of contamination of our sample is likely to be by
UCHII regions. These can easily be identified through the free-free
emission emitted by their ionised nebular gas. Radio continuum
observations are thus an essential part of our follow-up
programme. We have completed a programme of 5~GHz observations using
the ATCA (\citealt{urquhart2007a}) and the VLA (Urquhart et al. in
prep.) of all MYSOs candidates. These observations have a spatial
resolution of 1\arcsec\ 
and a sensitivity of 1~mJy (3$\sigma$), sufficient to detect a B0.5 or
earlier type star on the far side of the Galaxy
(\citealt{giveon2005}). About 25\% of
RMS sources have detectable radio emission, the majority of
which are classified as UCHII regions, due to their morphologies and tight correlation around the
Galactic mid-plane.

Mid-infrared imaging has been used to identify genuine point sources
from more extended emission. Given the typical size of MYSOs at
10~\mum\ is 0.005~pc (\citealt{churchwell1990}) and a spatial
resolution of 1\arcsec\ we would not expect to be able to resolve a
MYSO at distances $>$ 1~kpc. However, UCHII regions with a typical
size of $\sim$0.1~pc would be resolved at a distance of 15~kpc
($\sim$1.4\arcsec) as would most PNe. Therefore, mid-infrared imaging can
prove extremely useful in identifying, and removing extended
mid-infrared objects. Moreover, mid-infrared imaging is complementary to the
radio continuum observations and can be used to obtain accurate
astrometry and avoid excluding MYSOs that are located in close
proximity to UCHII regions. For the majority of our sources the
GLIMPSE data have been used for this purpose, however, the $\sim$700 sources located
outside the GLIMPSE region have been observed using UKIRT,
Gemini and the ESO 3.6~m telescope at La Silla
(\citealt{mottram2007}).

To obtain kinematic distances we have made molecular line observations
towards every source using $^{13}$CO ($J$=2--1) and ($J$=1--0)
transitions and Mopra, Onsala and Purple Mountain Observatory
telescopes, and the JCMT (\citealt{urquhart2007b}; Urquhart et al. submitted to A\&A). We have complemented these observations with archival data
extracted from the Galactic Ring Survey (GRS; see
\citealt{jackson2006} for
details). In cases where
multiple CO components are detected towards an RMS source we have used a
combination of nearby methanol and water masers, or CS ($J$=2--1)
observations to identify the component associated with the MYSO
candidate. We have calculated kinematic distances using the source
\vlsr\ and the rotation curve of \citet{brand1993}; these are crucial
for calculating luminosities and thus distinguishing between nearby
low- and intermediate-mass YSOs from genuine MYSOs. The majority
($\sim$80\%) of our sample is located within the solar circle and
therefore the rotation curve produces two possible kinematic
distances, referred to as the `near' and `far' distances; this is
known as the kinematic distance ambiguity. We have used the HI
self-absorption technique (\citealt{jackson2002}) and HI data from the
International Galactic Plane Survey (IGPS; for details see
http://www.ras.ucalgary.ca/IGPS/) to resolve these distance
ambiguities (e.g., \citealt{busfield2006}).

While the observations just discussed are useful in eliminating the
vast majority of UCHII regions, PNe and low- and intermediate-mass
YSOs, they are not so effective at eliminating evolved stars. We have
eliminated most of these using near infrared images in
combination with other available data since they tend to appear as
isolated point sources at all wavelengths, and are not generally
associated with strong CO emission ($^{12}$CO ($J$=1--0 and $J$=2--1)
typically less than 1~K; \citealt{loup1993}). In contrast, near-IR
images of MYSOs often show associated shocked emission from outflows,
reflection nebulosity, extinction lanes, as well as clusters of lower
mass YSOs.

The combination of near- and mid-infrared imaging, radio continuum and
molecular line observations allows us to eliminate most of the
contaminating sources. However, there are a few weak UCHII
regions, or chance alignments of evolved stars with molecular
material, that are not identified. To weed out any remaining contaminating
sources, and to confirm the identifications of our final sample of
MYSOs, we are obtaining near-infrared spectroscopy from which a
definitive classification can be made (e.g., \citealt{clarke2006}). We
have obtained spectra towards approximately 350 RMS sources so far.

\begin{figure}
\begin{center}

\includegraphics[width=0.59\textwidth]{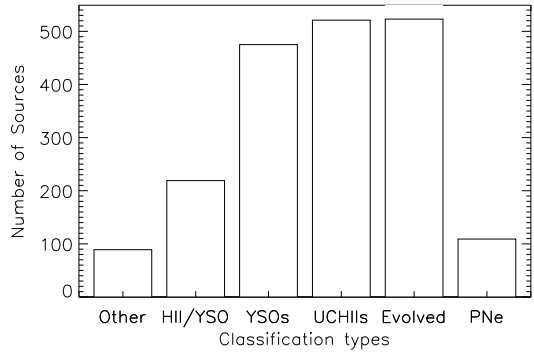}

\caption{\label{fig:classification} Classification and distribution of
RMS sources identified so far using data obtained from our
multi-wavelength observations.}

\end{center}
\end{figure}

\section{Results}

As the observational data have become available they have been used to
classify the sources in our sample. So far we have classified
all but a few of the $\sim$2000 sources and have identified $\sim$450 YSOs and a
further $\sim$500 UCHII regions (see Fig.~\ref{fig:classification} for
a detailed breakdown). Along with the YSOs and UCHII regions so far
identified, a further $\sim$200 sources have been identified as either
YSOs and/or UCHII regions. We are waiting on additional data
before making a final classification. Potentially, taking into account
these sources the final number of YSOs and UCHII regions identified so far will be significantly larger.

\begin{figure}[!h]
\begin{center}
\includegraphics[width=0.49\textwidth]{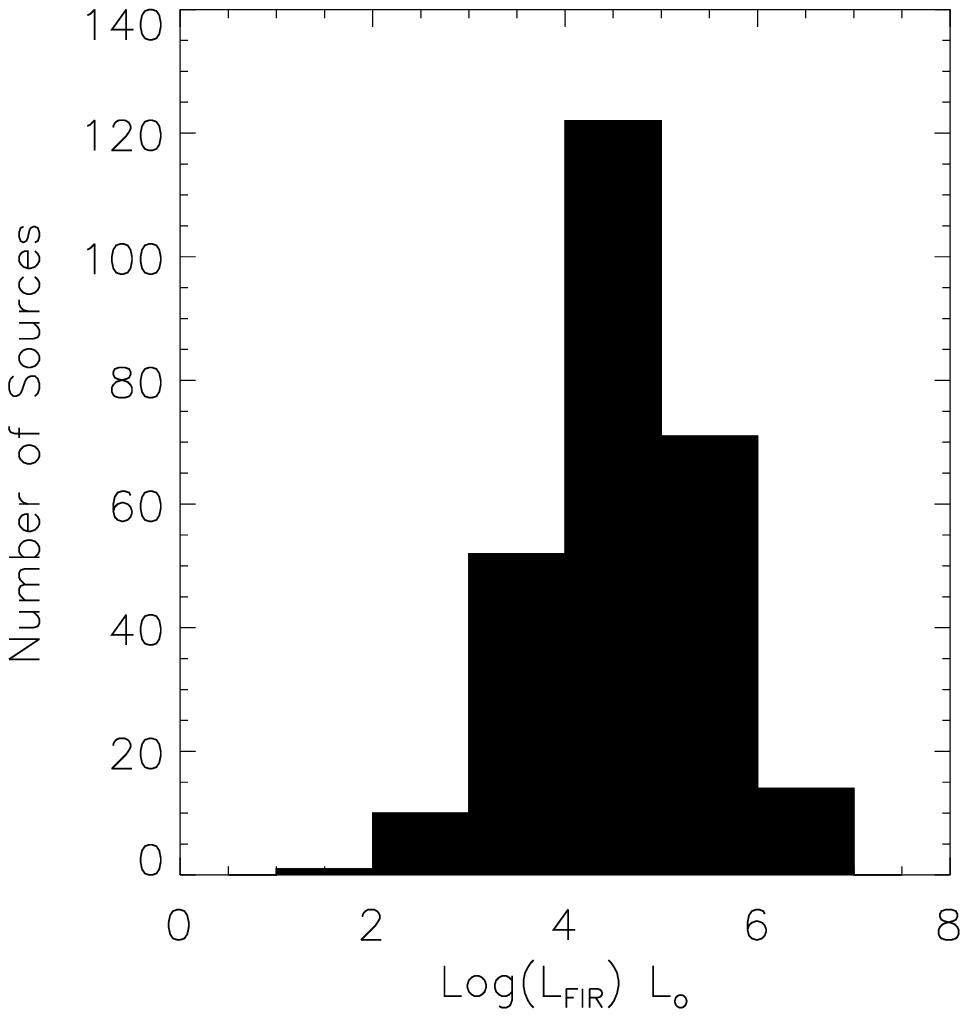}
\includegraphics[width=0.49\textwidth]{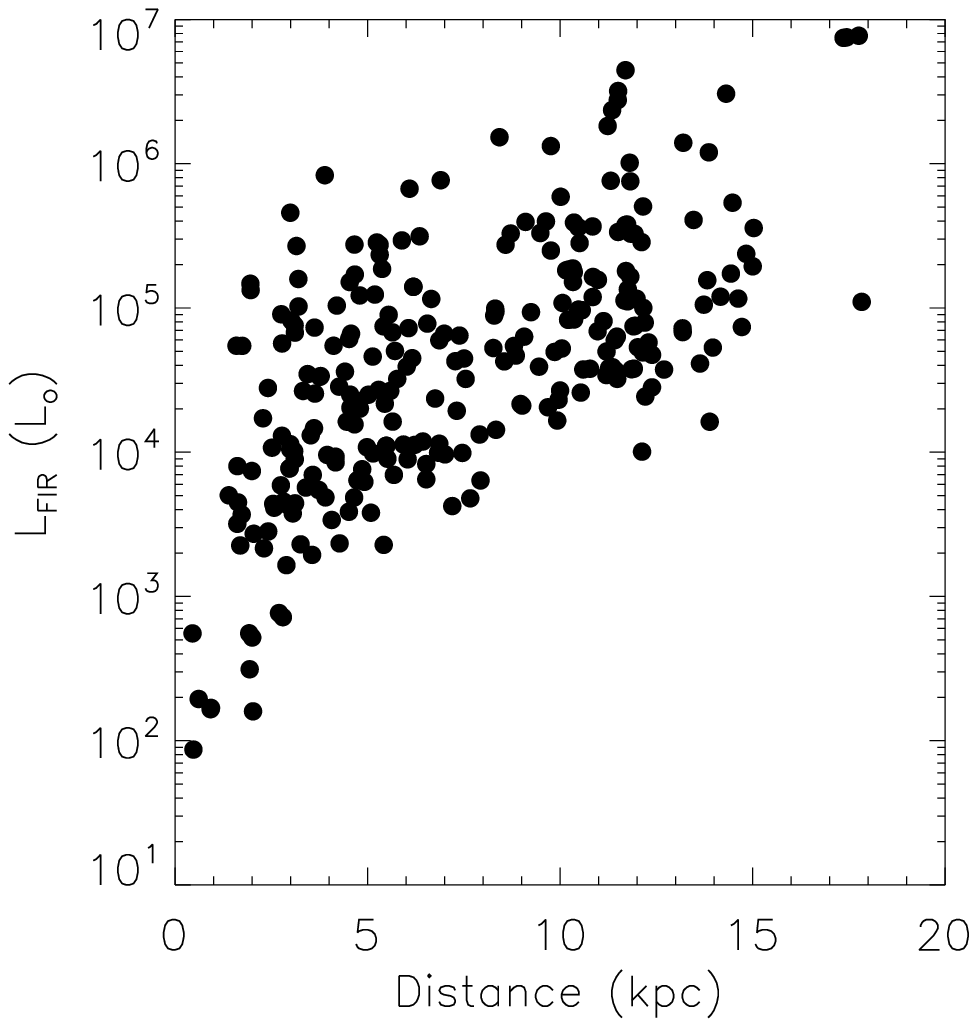}
\caption{\label{fig:luminosity}Left panel: Histogram of the luminosities for sources identified as either YSOs or UCHII regions and for which good IRAS data was available. These luminosities have been calculated following \citet{emerson1988}. Right panel: Distribution of source luminosities as a function of heliocentric distance.}

\end{center}
\end{figure}

In the left panel of Fig.~\ref{fig:luminosity} we present a histogram of
the luminosities of sources that have been classified as either YSOs
or UCHII regions and for which good quality IRAS data are available
($\sim$25\%), and for which any distance ambiguities have been
resolved. In the right panel of Fig.~\ref{fig:luminosity} we plot
the distribution of luminosities as a function of heliocentric
distance. These plots reveal that a number of the sources identified
as YSOs are actually nearby low- and intermediate-mass YSOs. However,
the vast majority have luminosities consistent with late O and early B-type stars.

\begin{figure}[!h]
\begin{center}

\includegraphics[width=0.99\textwidth]{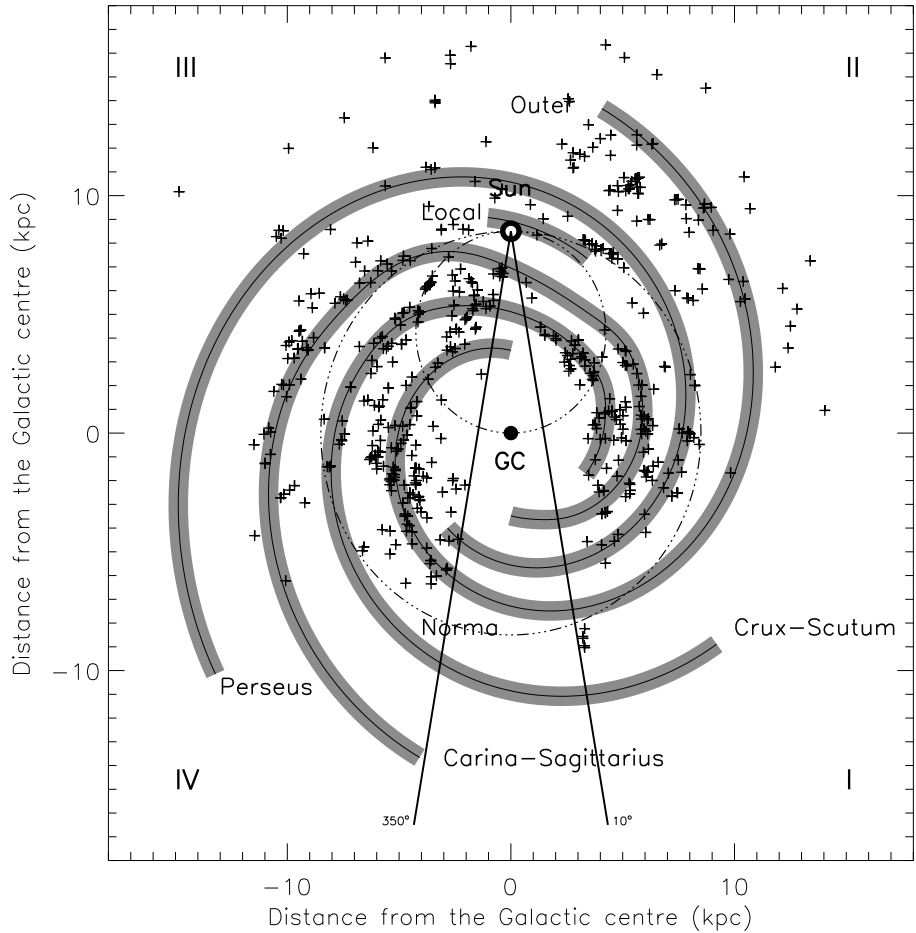}\\
\end{center}

\caption{\label{fig:gal_rms_distribution}Galactic distribution of RMS
sources identified as either UCHII regions or YSOs. The location of the spiral arms
taken from model by \citet{taylor1993} and updated by
\citet{cordes2004} are over-plotted in grey. The positions of the
Galactic centre and Sun are shown by black filled and unfilled
circles respectively. The Roman numerals in the corners refer to the
Galactic quadrants and the two thick black lines which originate from
the location of the Sun enclose the region excluded from our
survey. The two dashed circles indicate the solar circle and tangent
points.}

\end{figure}

We have determined kinematic distances for all YSOs and UCHII regions,
for which an unambiguous \vlsr\ could be attributed, using the rotation
curve of \citet{brand1993}. Using these distances and source positions
in Fig.~\ref{fig:gal_rms_distribution} we plot the distribution of
YSOs and UCHII regions on a face-on view of the Galactic plane. On
this plot we have indicated the positions of the Galactic centre, the
Sun, the solar circle and the spiral arms (see figure caption for
details). Massive stars are thought to be primarily associated with
the spiral arms, and thus could potentially be used as a probe of
Galactic structure. Comparing the distribution of our sources and the
position of the spiral arms (from the model by \citealt{taylor1993} and
updated by \citealt{cordes2004}) we find them to be reasonably well
correlated with each other, especially taking into account the errors
in the kinematic distances (typically $\pm$1~kpc). The correlation is
particularly apparent towards the inner Galactic arms (i.e., 1st and
4th quadrants), but quite poor for sources in the outer Galaxy.

\section{Summary}

The RMS survey aims to produce a large, unbiased sample of massive
young stellar objects (MYSOs) located throughout the Galaxy. We have
colour-selected approximately 2000 MYSO candidates from the MSX and
2MASS point source catalogues that have colours similar to known
MYSOs. We have almost completed a multi-wavelength programme of
follow-up observations to distinguish between genuine MYSOs and types
of embedded or dust enshrouded objects that have similar mid-infrared
colours that contaminate our sample. So far using these observations
we have unambiguously classified $\sim$90\% of our sample, approximately half of
which are identified as either YSOs and UCHII regions. The
luminosities of a sub-sample of these sources, for which good IRAS
data are available, confirms that we are detecting a significant number
of genuine MYSOs. When complete, we estimate our database will contain
a well selected sample of $\sim$500 bona fide MYSOs and a further
500--600 UCHII regions, along with complementary multi-wavelength
observations.

Now the classification is almost complete we have begun a number of
observational programmes to exploit our sample and to address many
open questions concerned with massive star formation. These include:
high resolution observations of potentially triggered regions to
investigate different triggering mechanisms; chemical surveys using
millimetre lines and IR absorption features to investigate the
chemistry and possible chemical evolution of MYSOs; investigation of
molecular outflows associated with MYSOs; and high resolution spectroscopy of accretion disks. All of these programmes are
being conducted on sub-samples selected as a function of luminosity,
distance and location. The whole sample and all of the observational
results of our multi-wavelength campaign are available at
www.ast.leeds.ac.uk/RMS.

\acknowledgements %%% Text of acknowledgements runs on after this command.

We would like to thank the directors and staff of the various telescopes mentioned in this article for their help and support during our observations. JSU is supported by a postdoctoral grant from the Science and Technology Facilities Council (STFC).

%%% THE BIBLIOGRAPHY
%%%
%%% CONSULT SECTION 3 OF "INSTRUCTIONS FOR AUTHORS" FOR HOW TO USE NATBIB.
%%% AUTHORS ARE ENCOURAGED TO USE EITHER THE "THEBIBLIOGRAPY" ENVIRONMENT
%%% BY UNCOMMENTING (DELETING THE "%" SYMBOL) THE COMMANDS BELOW, OR BY
%%% USING THE BIBTEX ENVIRONMENT. TO FIND OUT WHICH IS APPLICABLE TO YOUR
%%% CONTRIBUTION, CONSULT THE VOLUME EDITORS FOR YOUR PROCEEDINGS.
%%%

\end{document}